\documentclass{aa}  

\usepackage{graphicx}
\usepackage{txfonts}
\usepackage{xspace}
\usepackage[range-units=single,range-phrase=\text{--}]{siunitx}
\usepackage{natbib}
\usepackage{makecell}
\usepackage[switch]{lineno}
\usepackage[colorlinks,allcolors=blue]{hyperref}
\usepackage{changes}
\DeclareSIUnit\pc{pc}
\DeclareSIUnit\kpc{kpc}
\DeclareSIUnit\erg{erg}  
\DeclareSIUnit\year{yr}
\DeclareSIUnit\TeV{TeV}
\DeclareSIUnit\GeV{GeV}
\DeclareSIUnit\PeV{PeV}
\DeclareSIUnit\gauss{G}
\DeclareSIUnit\mas{mas}
\begin{document} 

   \title{Deriving pulsar pair-production multiplicities from pulsar wind nebulae using H.E.S.S. and LHAASO observations}

   \author{ S.T. Spencer\inst{1,2} \and A.M.W. Mitchell\inst{1}}

   \institute{Erlangen Centre for Astroparticle Physics (ECAP),  Friedrich-Alexander-Universität Erlangen-Nürnberg, Nikolaus-Fiebiger-Str. 2, 91058 Erlangen, Germany\\ \email{samuel.spencer@fau.de}
   \and
    Department of Physics, Clarendon Laboratory, Parks Road, Oxford, OX1 3PU, United Kingdom
             }

   \date{\today}
    \titlerunning{Pulsar pair production multiplicities from pulsar wind nebulae}
    \authorrunning{Spencer and Mitchell}
  \abstract
   {Pulsar wind nebulae (PWNe) dominate the galactic gamma-ray sky at very high energies and they are major contributors to the leptonic cosmic ray flux. However, the question of whether or not pulsars also accelerate ions to comparable energies has not yet been experimentally confirmed.}
   {We aim to constrain the birth period and pair-production multiplicity for a set of pulsars. In doing so, we aim to constrain the proportion of ions in the pulsar magnetosphere and, hence, the proportion of ions that could enter the pulsar wind.}
   {We estimated possible ranges of the value of the average pair production multiplicity for a sample of 26 pulsars in the Australia Telescope National Facility (ATNF) catalogue, which have also been observed by the High Energy Stereoscopic System (H.E.S.S.) telescopes. We then derived lower limits for the pulsar birth periods and average pair production multiplicities for a subset of these sources where the extent of the pulsar wind nebula and surrounding supernova shell have been measured in the radio. We also derived curves for the average pair production multiplicities as a function of birth period for sources recently observed by the Large High Altitude Air Shower Observatory (LHAASO).}
   {We show that there is a potential for hadrons entering the pulsar wind for most of the H.E.S.S. and LHAASO sources we consider here, which is dependent upon the efficiency of luminosity conversion into particles. We also present estimates of the pulsar birth period for six of these sources,  all falling into the range of $\simeq$10-50\,ms.}
   {}


   \maketitle
%

\section{Introduction}
\label{sec:intro}

Pulsar wind nebulae (PWNe) follow a supernova explosion and the pulsar's ultra-relativistic wind flows into the surrounding medium, producing a wind termination shock that can accelerate particles to relativistic energies \citep{gaensler_slane_review,crab_pwn_kargaltsev_review}. We know that PWNe are the most numerous class of galactic very-high-energy gamma-ray emitters (representing  $\sim$40\% of known galactic sources) and have been observed across the electromagnetic spectrum for $\simeq$50 years \citep{2008ICRC....3.1341W,pwn_pop_paper}. A key question regarding PWNe is whether the gamma-ray emission produced is solely a result of inverse Compton scattering of relativistic electrons on background fields or if pion production caused by protons originating from the pulsar could contribute to the emission at the highest energies \citep{bp97,bednarek2003}.

A key metric that governs whether hadrons can likely escape the pulsar surface into the PWN is the average pair production multiplicity $\langle\kappa\rangle$. This describes the number of e+/e- pairs that escape the pulsar light cylinder per electron that escapes from the pulsar's surface (as a result of cascades produced by gamma-ray Bremsstrahlung). This places some constraints on potential hadrons originating from the pulsar, as such particles do not multiply in cascades in the pulsar magnetosphere; thus, they can only make up a fraction of $1/\langle\kappa\rangle$ of the total particles at most \citep{Kirk}. Therefore for sources with a low pair-production multiplicity there exists the possibility of hadrons escaping into the wind. In this paper, we aim to determine the value of $\langle\kappa\rangle$ for a number of gamma-ray-emitting PWNe as observed by the High Energy Stereoscopic System (H.E.S.S.) gamma-ray telescopes \citep{hess}. This represents roughly a quarter of known PWN \citep{2012AdSpR..49.1313F,Giacinti_2020}. We also make use of radio observations, as these can help to constrain the size of the PWN and surrounding supernova remnant (SNR), even once the associated gamma-ray emission has faded. On this basis, we can also place constraints on the pulsar birth period for a subset of these systems. We  also set constraints on the potential values of $\langle\kappa\rangle$ for a number of sources recently observed by the Large High Altitude Air Shower Observatory \citep[LHAASO;][]{cao2022large}. This paper builds upon the previous work by \citet{de_Jager_2007}, who introduced the concept of deriving pair-production multiplicity constraints from gamma-ray data and comparing them to estimates of the pulsar birth period obtained from radio data; however, we utilise a greater number of considered sources, updated observations, and updated modelling.

This paper is organised as follows: In Sect. \ref{sec:modelling}, we introduce the theoretical background to our models and discuss the input parameters to them. In Sect. \ref{sec:results}, we present our main results. In Sect. \ref{sec:discussion}, we discuss the implications for cosmic ray production, the effect of varying free parameters in the models, and the prospects for future observations. Finally, in Sect. \ref{sec:conclusions} we present our conclusions.
\section{Modelling theory}
\label{sec:modelling}
\subsection{Deriving $P_0$}
\label{sec:p0}
Following the reasoning of \citet{van_der_Swaluw_2001}, the initial spin period of a pulsar embedded in a PWN within a larger SNR can be determined for systems aged $\sim 10^3-10^4$\,yr using the relation
\begin{equation}
P_0 = 2 \pi \left[\frac{2E_0}{\eta_1 I}\left(\frac{R_{PWN}}{\eta_3 R_{SNR}} \right)^3 + \left(\frac{2\pi}{P_t} \right)^2  \right]^{-1/2},
\label{eq:P0}
\end{equation}
where $E_0$ is the total mechanical energy of the SNR which we take to be $10^{51}\,\mathrm{erg}$, $I\approx1.4\times10^{45}\,\mathrm{g\,cm^2}$ is the moment of inertia of the neutron star \citep{MOLNVIK1985239}, $R_{SNR}$ and $R_{PWN}$ are the SNR and PWN radii as measured in the radio (as these mark the `true' extent of the PWN; \citealt{van_der_Swaluw_2001}), and $P_t$ is the period of the pulsar at the present time. Then, $\eta_1$ and $\eta_3$ are dimensionless scaling parameters that relate the radius of the PWN $R_{PWN}$ to the ratio of the SNR $R_{SNR}$:
\begin{equation}
R_{PWN}(t)=\eta_3(t)(\eta_1 E_{SD}/E_0)^{1/3}R_{SNR}(t).
\label{eq:RPWN}
\end{equation}
Here, $E_{SD}$ is the total spin-down energy of the pulsar, while $\eta_1$ effectively a proxy for the strength of synchrotron losses, neglecting these and setting $\eta_1=1$ and $\eta_3=1.02$ is suggested by \citet{van_der_Swaluw_2001}, leading to a systematic underestimation of the initial spin rate which propagates through to our estimates of $\langle\kappa\rangle$. Equations \ref{eq:P0} and \ref{eq:RPWN} are only valid in the subsonic phase of the PWN's evolution, after most of the energy in the pulsar wind has been deposited into the PWN, and follow the assumption that the PWN+SNR system is spherically symmetric \citep{van_der_Swaluw_2001}. \citet{de_Jager_2007} also considered the work of \citet{van_der_Swaluw_2001}, although they ultimately adopted a fixed $P_0$ in their multiplicity analysis as a constraint of radio observations available at the time. We utilised a $P_0$ value calculated for each H.E.S.S. source for which we now have radio data in our analysis.
\subsection{Determining $N_{el}$ and $\langle\kappa\rangle$}
The number of Goldreich-Julian electrons at the current time, $t$, which represents the total number of electrons that have been stripped from the polar caps of the pulsar, is given by the integral
\citep{1969ApJ...157..869G}:\begin{equation}
N_{GJ}=\int_{t=0}^{t=-\tau(P_0)} \frac{[6c\dot{E}(t)]^{1/2}}{e} (-dt)\,
\label{eq:gj}
.\end{equation}
 To calculate $\dot{E}$, we took values of $\dot{E}(t)$ today, as calculated in \citet{Giacinti_2020}, and assumed that the energy output of the pulsars evolves as 
\begin{equation}
\dot{E}(t)=\dot{E}_0\left(1+\frac{t}{\tau_0}\right)^{-\alpha},
\end{equation}
with $\dot{E}_t$ the value of $\dot{E}$ at the current time, the constant $\tau_0$ set to $10^3\,$years and $\alpha=(n+1)/(n-1)$ assumed to be equal to 2 for a braking index $n=3$. $N_{GJ}$ is then integrated numerically.
The average pair production multiplicity is then \citep{de_Jager_2007}:
\begin{equation}
\langle \kappa \rangle = \frac{N_{el}}{2N_{GJ}}
\label{eq:kappa}.
\end{equation}

We estimate the number of electrons in the PWNe observed by H.E.S.S. \citep{2005A&A...432L..25A,pwn_pop_paper} by following the approach of \citep{Giacinti_2020} and approximating the electron spectrum as a broken power law with a single spectral parameter (BPL1) such that
\begin{eqnarray}
N_{el}=E_{tot}\left(\frac{(2-\Gamma)E_0^{1-\Gamma}}{E_2^{2-\Gamma}-E_1^{2-\Gamma}}\right) \times \left(\frac{(E_2/E_0)^{1-\Gamma}-(E_1/E_0)^{1-\Gamma}}{1-\Gamma}\right)
\label{eq:nel}
,\end{eqnarray}
where $\Gamma=2.2$, $E_0$ is assumed to be $0.1\,\mathrm{TeV}$, $E_1$ is the energy threshold of the gamma-ray observations (taken either from \citealt{HGPS_paper} or \citealt{2005A&A...432L..25A} for G0.9+0.1/PSR J1747-2809) and $E_2$ is assumed to be $10\,\mathrm{TeV}$. This is because the radiative lifetime of electrons above this energy is likely to be less than the typical ages of pulsars in our sample, since this scales with electron energy, $E_e$, as $\sim 10^4 (B/10\,\mathrm{\mu G})^{-2}(E_e/10\,\mathrm{TeV})^{-1}\,\mathrm{yr}$ \citep{Giacinti_2020}.
The values for the total energy in electrons $E_{tot}$ are taken from the modelling in \citet{Giacinti_2020}. We can then calculate the age, $\tau$, as a function of $P_t$, the pulsar period today, the period derivative today, $\dot{P}_t$, the initial period, $P_0$, and the braking index, $n$, such that \citep{gaensler_slane_review}:
\begin{equation}
\tau(P_t,\dot{P}_t,P_0,n)=\left(1-\left(\frac{P_0}{P_t}\right)^{n-1}\right) \times \frac{P_t}{(n-1)\dot{P}_t}. 
\label{eq:tau}
\end{equation}

Given the dependence of Eq. \ref{eq:gj} upon $\tau(P_0)$ and therefore $P_0$ by Eq. \ref{eq:tau}, we can find curves of $\langle\kappa\rangle$ as a function of $P_0$ (where $P_0$ is constrained to be shorter than the measured $P_t$ today). As these curves do not consider electrons producing emission outside the gamma-ray range, these curves serve as strict lower limits for the true pulsar multiplicity. This technique to determine these curves was first presented in \citep{de_Jager_2007}. As we can simultaneously find $P_0$ through the ratio of $R_{PWN}/R_{SNR}$ and Eq. \ref{eq:P0}, for sources where $P_t,R_{PWN},R_{SNR}$ are known and $N_{el}$ can be modelled, one can estimate a specific value for the lower bound of the pair production multiplicity. This is found by finding the intersection between the estimate for $P_0$ using Eq. \ref{eq:RPWN}, and the curves of $\langle\kappa\rangle(P_0)$ as estimated using Eqs. \ref{eq:gj}, \ref{eq:nel}, and \ref{eq:tau}\footnote{Jupyter notebooks showing this analysis can be found at \url{www.github.com/STSpencer/psrprops}.}.
\begin{table*}[!htbp]
\centering
\caption{Input parameters to our modelling and the corresponding spectral model used. }
\begin{tabular}{c c c c c c c c c c c c c}
\hline
\hline
\makecell{ATNF Name\\} & \makecell{$R_{SNR}$ \\$\mathrm{[pc]}$}& \makecell{$R_{PWN}$ \\$\mathrm{[pc]}$} & \makecell{$P_t$ \\$\mathrm{[ms]}$} & \makecell{$\dot{P}_t$ \\ $\mathrm{[\times 10^{-13}\ s/s]}$} & \makecell{Model\\} & \makecell{$E_0$ \\$\mathrm{[TeV]}$} & \makecell{$E_1$\\ $\mathrm{[TeV]}$}& \makecell{$E_{cut}$\\ $\mathrm{[TeV]}$} &\makecell{$E_2$\\ $\mathrm{[TeV]}$} & \makecell{$\Gamma$\\} & \makecell{$\Gamma_2$\\}&\makecell{Refs.}\\
\hline
J1833-1034 & 2.98 & 0.8 & 61.8 & 2.02 & BPL1 & 0.1 & 0.39 & - & 10 & 2.2 & -&$1,2$\\
J1513-5908 & 38.4 & 19.2 & 151.6  & 15.3 & BPL1 & 0.1 &  0.61 &-& 10 & 2.2 & -&$1,2$\\
J1930+1852 & 10.8 & 2.7 & 136.9 & 7.50 & BPL1 & 0.1 & 0.89 & - & 10 & 2.2 & -&$1,2$\\
J1846-0258 & 2.6 & 0.58 & 326.6 & 71.1& BPL1 & 0.1 & 0.40 & - & 10 & 2.2 & - &$1,2$\\
J0835-4510 & 19.5 & 12.2 & 89.3& 1.25 & BPL1& 0.1 & 0.61 & - & 10 & 2.2 & -&$1,2$\\
J1747-2809 & 19.8 & 2.5 & 52.2 & 1.56 & BPL1& 0.1 & 0.17 & - & 10 & 2.2 & -&$1,3$\\
J2021+3651 & - & - & 103.7 &0.957& PLEC & 1& 25 & 900 & 1400 & 1.4 & -&$4$\\
J1841-0345 & - & - & 112.9 &1.55 & PLEC & $1\times 10^{-5}$ &7.0& 72 & 740& 2.2 & -& $5$\\
J1849-0001  & - & - & 38.5 & 0.142 & PL & 0.5 & 10 &- & 100 &2.5 & -&$6$\\
J1826-1334  & - & - & 101.5 & 0.753 & BPL2 & 0.7 & 0.9 & - & 42 & 1.4 & 3.25& $7$\\
\hline
\end{tabular}
\tablefoot{$R_{SNR}$ and $R_{PWN}$ values are taken from \citet{Giacinti_2020} and references therein. $P_t$ and $\dot{P}_t$ values are taken from the ATNF catalogue \citep{Manchester05}. The possible spectral models are described in Sect. \ref{sec:modelling}, note that these models assume different radiation fields and should not be considered equivalent for all sources.}
\tablebib{(1) \citet{Giacinti_2020}, (2) \citet{HGPS_paper}, (3) \citet{2005A&A...432L..25A}, (4) \citet{2023ApJ...954....9W}, (5) \citet{hawc1844}, (6) \citet{Amenomori_2023}, (7) \citet{2019A&A...621A.116H} }
\label{table:inputtab}
\end{table*}
We also derived pair-production multiplicity curves for those sources observed by LHAASO with published results of modelling of their gamma-ray emission. We note that these publications make different assumptions about, for example, the associated radiation fields compared to the H.E.S.S. sources. For the Dragonfly Nebula, associated with PSR\,J2021+3651, we utilised a power law model with an exponential cut-off (PLEC), namely,
\begin{equation}
N_{el}\propto \int_{E_0}^{E_2} \left(\frac{E}{E_1}\right)^{-\Gamma}\exp\left(\frac{E}{E_{cut}}\right) dE
,\end{equation}
following \citet{2023ApJ...954....9W}. Here, we use a total energy of $3.9\times10^{48}\,\mathrm{erg}$ to determine the proportionality constant numerically, a reference energy of $E_1=\mathrm{1\,TeV}$, a lower energy threshold of $E_0=\mathrm{25\,TeV}$, a cutoff of $E_{cut}=\mathrm{900\,TeV}$, and a maximum energy of $E_2=1400\,\mathrm{TeV}$ following \citet{cao2023lhaaso}, along with a spectral index of $\Gamma=\mathrm{1.4}$. For PSR J1841-0345, we also followed \citet{hawc1844} by using a power-law spectrum with an exponential cutoff, with a proportionality constant of $7.1\times 10^{31}\,\mathrm{erg}$, a low-energy threshold of $E_0=\mathrm{10\,MeV}$, a reference energy of $E_1=\mathrm{7\,TeV}$, a cut-off energy of $E_{cut}=\mathrm{72\,TeV}$, a maximum energy of $E_2=\mathrm{740\,TeV,}$ and a spectral index of $\Gamma=\mathrm{2.2}$. For PSR\,J1849-0001, we followed \citet{Amenomori_2023}, with the power-law model expressed as\begin{equation}
N_{el}\propto \left[\frac{1}{1-\Gamma}\left(\frac{E_2}{E_0}\right)^{1-\Gamma}-\frac{1}{1-\Gamma}\left(\frac{E_1}{E_0}\right)^{1-\Gamma}\right]
.\end{equation}
Here, we have a low energy threshold of $E_0=\mathrm{0.5\,TeV}$, a reference energy of $E_1=\mathrm{10\,TeV}$,  a high-energy threshold of $E_2=\mathrm{100\,TeV,}$ a spectral index of $\Gamma=2.5,$ and a total energy of $2.8\times10^{47}\,\mathrm{erg}$ (from which, the proportionality constant can  be derived numerically). For PSR J1826-1334, we follow \citet{2019A&A...621A.116H} in using a broken power-law model with a two spectral parameters (BPL2), such that
\begin{equation}
N_{el}\propto \left[\int_{E_0}^{E_1}E^{-\Gamma} dE+\int_{E_1}^{E_2}E^{-\Gamma_2} dE \right]
.\end{equation}
Here, we have a low-energy threshold of $E_0=\mathrm{0.7\,TeV}$ \citep{HGPS_paper}, a break energy of $E_1=\mathrm{0.9\,TeV}$, a maximum energy of $E_2=\mathrm{42\,TeV,}$ a lower-energy spectral index of $\Gamma=1.4,$ and a high-energy spectral index of $\Gamma_2=3.25,$ with a total energy of $5.5\times 10^{48}\,\mathrm{erg}$ (from which, the proportionality constant can be determined numerically). A summary of the input parameters to our modelling, along with the spectral models used for each source, are shown in Table \ref{table:inputtab}. We obtained values for $R_{SNR}$ and $R_{PWN}$ from \citep{Giacinti_2020} and references therein.
\section{Results}
\label{sec:results}
\begin{figure}
\centering
    \includegraphics[width=\columnwidth]{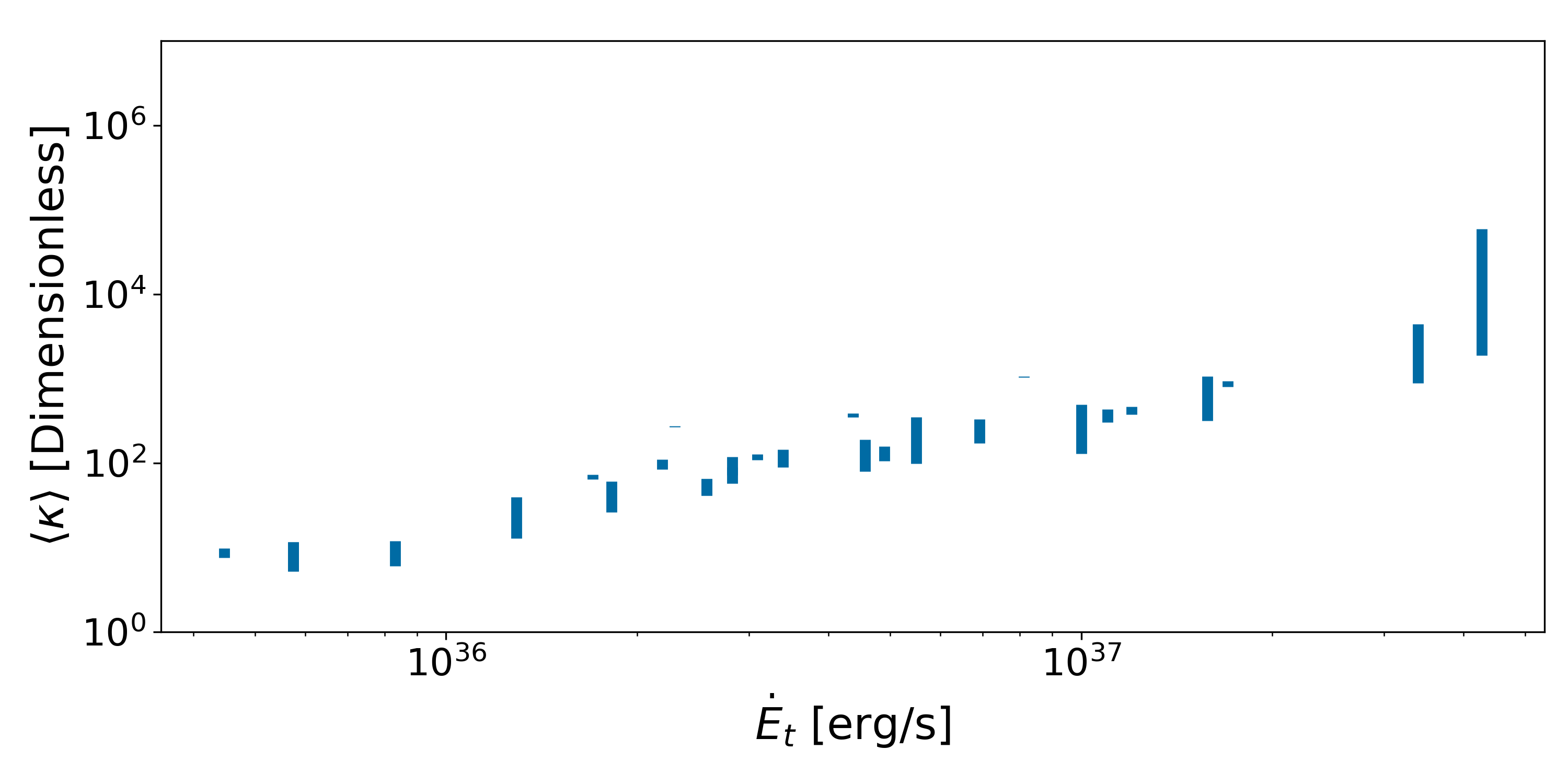}
    \caption{Maximum and minimum values of the lower limit for $\langle\kappa\rangle$ for the pulsars for which we can estimate $N_{el}$. The minimum bounds for this lower limit correspond to assuming a birth period of 10\,ms, and the maximum corresponds to assuming a birth period of 50\,ms.}
    \label{fig:kapparange}
\end{figure}

For 26 pulsars in the Australia Telescope National Facility ATNF catalogue, whose regions have been observed by H.E.S.S., we can place constraints on the lower limit of $\langle\kappa\rangle$. We do this by assuming that the birth period is in the range 10-50\,ms, based on the reasoning that pulsars with birth periods below 10\,ms are unlikely to exist \citep{10.1111/j.1365-2966.2007.12821.x} and PWNe associated with longer period pulsars are unlikely to produce gamma-ray emission. The resulting ranges for the possible lower-limit of $\langle\kappa\rangle$ are shown in Fig. \ref{fig:kapparange}, with numerical values provided in Table \ref{table:a1}.
\begin{table}[!htbp]
\centering
\caption{Estimates of the maximum and minimum values of the lower limit for $\langle\kappa\rangle$ for the H.E.S.S./ATNF sample.}
\label{table:a1}
\resizebox{\columnwidth}{!}{%
\begin{tabular}{c c c c}
\hline
\hline
\makecell{ATNF Name\\} & \makecell{$\dot{E}_t$\\$\mathrm{[\times 10^{36} erg/s]}$} & \makecell{$\langle\kappa\rangle_{min}$\\$\mathrm{[Dimensionless]}$} & \makecell{$\langle\kappa\rangle_{max}$\\$\mathrm{[Dimensionless]}$} \\
\hline
J1833-1034 & 33.9 & 884 & 4430 \\
J1513-5908 & 17.0 & 802 & 938 \\
J1930+1852 & 12.0 & 374 & 464 \\
J1420-6048 & 10.0 & 130 & 495 \\
J1846-0258 & 8.13 & 1030 & 1060 \\
J0835-4510 & 6.92 & 173 & 333 \\
J1838-0655 & 5.50 & 98 & 349 \\
J1418-6058 & 4.90 & 107 & 159 \\
J1357-6429 & 3.09 & 109 & 128 \\
J1826-1334 & 2.82 & 72 & 119 \\
J1119-6127 & 2.29 & 268 & 273 \\
J1301-6305 & 1.70 & 64 & 73 \\
J1747-2809 & 42.7 & 1880 & 59185 \\
J1617-5055 & 15.8 & 316 & 1072 \\
J1023-5746 & 11.0 & 304 & 435 \\
J1856+0245 & 4.57 & 80 & 191 \\
J1640-4631 & 4.37 & 351 & 386 \\
J1709-4429 & 3.39 & 89 & 145 \\
J1907+0602 & 2.82 & 57 & 90 \\
J1016-5857 & 2.57 & 42 & 65 \\
J1803-2137 & 2.19 & 85 & 111 \\
J1809-1917 & 1.82 & 26 & 61 \\
J1718-3825 & 1.29 & 13 & 39 \\
J1028-5819 & 0.832 & 6 & 12 \\
J1833-0827 & 0.575 & 5 & 12 \\
J1857+0143 & 0.447 & 8 & 10 \\
\hline
\end{tabular}
}
\tablefoot{$\langle\kappa\rangle_{min}$ assumes a $P_0$ of 10\,ms and $\langle\kappa\rangle_{max}$ assumes a $P_0$ of 50\,ms. This corresponds to the entire vertical range as shown in Fig. \ref{fig:kapparange}.}
\end{table}
To determine whether hadrons originating from the pulsar could escape into the pulsar wind, we followed the approach of \citet{2015JCAP...08..026K}, who posited that ultra-high-energy cosmic rays can be produced by iron stripped from the pulsar surface being photo-dissociated by the radiation environment in the vicinity of the star. However, as we wish to make a more general statement about whether protons of $\sim$TeV-PeV energy could escape into the pulsar wind, we have made less extreme assumptions about the energy of the stripped iron nuclei; also, we used $P_0$ values derived from our modelling, rather than assuming $P_0<10\,\mathrm{ms}$. We derived a pair production multiplicity limit for each source $\langle\kappa\rangle_{\rm lim}$ using the formula \citep{2015JCAP...08..026K}:\ 
\begin{equation}
E_{\mathrm{CR}} \approx 1.2\times 10^{20} A_{56}\eta\langle\kappa\rangle_{\rm lim,4}I_{45}B_{13}^{-1}R_{\star,6}^{-3}\tau_{7.5}^{-1}\,\mathrm{eV}
\label{eq:ECR}
,\end{equation}where $E_{\mathrm{CR}}$ is the iron nuclei energy which we assume to be $3\,\mathrm{PeV}$, roughly equivalent to the cosmic ray `knee' \citep{Blasi_2013,annurev:/content/journals/10.1146/annurev-astro-082214-122457}, $A_{56}=1$ is the mass number of the stripped particles normalised to 56, and $B, R_{\star}, I$, and $\tau$ are the magnetic field strength, radius of the pulsar, moment of inertia, and age of the system, respectively (with subscripts $x$ representing normalisation to $10^x$ in cgs units). $I_{45}$ and $R_{\star,6}$ we assumed to be equal to 1, as they are non-trivial to determine \citep{_zel_2016}, whereas we calculated $\tau_{7.5}$  using Eq. \eqref{eq:tau} and $B_{13}$ was obtained using the equation: 
\begin{equation}
B=3.2\times10^{19}\sqrt{P_t\dot{P_t}}\,\mathrm{G}
\end{equation}
\citep{gaensler_slane_review}. Here, $\eta\leq1$ in Eq. \eqref{eq:ECR} is a luminosity conversion efficiency factor that is effectively unknown. Previous theoretical studies considering hadronic acceleration by pulsars have used values ranging from 0.1 to 1 \citep{kotera2014cosmic} and it could feasibly be of the order of 0.01 \citep{de_O_a_Wilhelmi_2022}, although \cite{de_O_a_Wilhelmi_2022} showed that for many known gamma-ray sources, it appears to be in the range 0.1 to 1 for particles downstream of the wind termination shock. As such, the conclusion reached for whether hadrons can escape into the pulsar wind for a given pulsar depends on the adopted value of $\eta$, as the intersection point between our derived $P_0$ value and the curve of $\langle\kappa\rangle(P_0)$ can fall above or below the derived value for $\langle\kappa\rangle_{\rm lim}$.

\begin{figure}
\centering
    \includegraphics[width=\columnwidth]{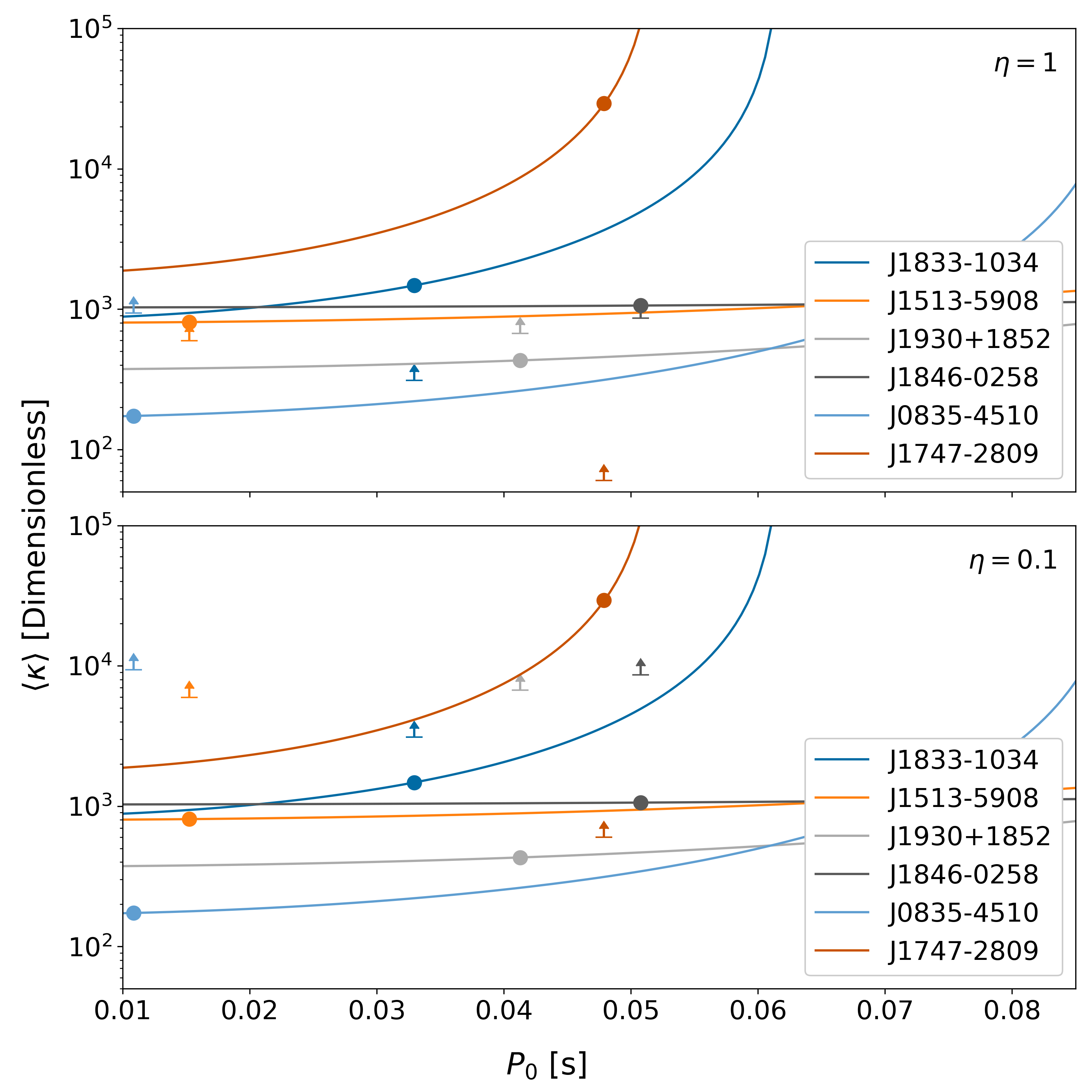}
    \caption{Intersections of our estimates for $P_0$ using Eq. \ref{eq:RPWN}, and the curves of $\langle\kappa\rangle(P_0)$ as estimated using Eqs. \ref{eq:gj}, \ref{eq:nel}, and \ref{eq:tau} (the latter is constrained such that $P_0<P_t$) are shown by circular markers. The intersection location represents the lower limit for the pair-production multiplicity (see Table \ref{table:kappatab}) for the associated values. Derived values for the pair-production multiplicity limit $\langle\kappa\rangle_{\rm lim}$ for each pulsar are shown as lower limit markers in matching colours to the multiplicity curve. The top panel assumes a luminosity conversion efficiency of $\eta=1$ in the determination of $\langle\kappa\rangle_{\rm lim}$, whereas the lower panel assumes $\eta=0.1$.}
    \label{fig:kappaplotdual}
\end{figure}

The derived initial periods, pair multiplicities and values for $\langle\kappa\rangle_{\rm lim}$ for the six H.E.S.S. sources, where we have both an estimate of $R_{PWN}/R_{SNR}$ and $N_{el}$ are shown in Fig. \ref{fig:kappaplotdual} for assumed values for $\eta$ of 1 and 0.1. The calculated values for $P_0$ and $\langle\kappa\rangle$ for these sources is summarised in Table \ref{table:kappatab}. We can confidently not exclude the possibility of hadronic escape into the wind for the pulsars J0835-4510 and J1930+1852, as they have intersection points between their $\langle\kappa\rangle(P_0)$ curve and the modelled value of $P_0$ below the marker representing $\langle\kappa\rangle_{\rm lim}$ in the maximal case of assuming $\eta=1$. However, we also cannot completely exclude this scenario for the other four H.E.S.S. pulsars depending on the $\eta$ value assumed. That said, for J1747-2809, the value of $\eta$ would have to be of the order of $10^{-3}$, so it is unlikely that hadrons escape into the wind for this source. 
The result that a higher efficiency factor for the production of hadrons at the pulsar corresponds to a greater likelihood of said hadrons escaping into the wind is expected, but our results quantify the effect of this for the first time. It was previously claimed in \citet{2015JCAP...08..026K} that there is a rough value of $\langle\kappa\rangle_{\rm lim}$ of $2m_p/m_e\approx3672$ for determining whether hadrons can escape into the wind, based on the specific hypothetical case of a young, $P_0\leq10\,\mathrm{ms}$ pulsar producing ultra-high-energy cosmic rays. This claim is is only likely to be correct to within one to two orders of magnitude. This is because (as seen from Fig. \ref{fig:kappaplotdual}) the constraints on hadronic escape reached for any particular source have that order uncertainty, depending on the measured or assumed values of $P$, $P_0$, and $\eta$. Despite the differences in our modelling, we obtained an order-of-magnitude similarity to the curve derived in \cite{de_Jager_2007} for PSR B1509-58 (PSR J1513-5908). For the Vela Pulsar (PSR J0835-4510), our lower limit curve is within the permitted range predicted in that work, however, the precise level of agreement depends on scaling to the magnetic field strength in the region. However, some discrepancies for the results for these two pulsars considered in both our work and \citet{de_Jager_2007} are to be expected, as we have based our results on more recent observational data and PWN modelling, as referenced in Table \ref{table:inputtab}. 

With the exception of PSR J1833-1034, we obtained birth periods that are consistent within a factor two of previous estimates. The birth period we derived for PSR J1833-1034 of 33.0\,ms is in stark contrast to the value of >55\,ms derived by \citet{2006ApJ...637..456C}; this is likely due to Eq. \ref{eq:P0} not being suitable for a source that is potentially of age $<1$\,kyr (much lower than its characteristic age of $\tau_c\approx 4.8$\,kyr), as suggested by \citet{Camilo_2009}. The validity of our pair-production multiplicity for this source is therefore also questionable. The birth period for PSR J1513-5908 we derive (15.2\,ms) is in good agreement with the value derived by \citet{Abdo_2010} of 16\,ms. Similarly for the Vela Pulsar PSR\,J0835-4510 the value of 10.9\, ms we obtain is in reasonable agreement with the value from \citet{Helfand_2001} of 6\,ms. The birth period of 47.9\,ms we obtain for PSR\,J1747-2809 is consistent with the prediction of \citet{Camilo_2009} who proposed it to be >40\,ms. It has been postulated that pulsars with approximately millisecond birth periods could explain the ultra-high energy all-particle cosmic ray flux if they exist \citep{2015JCAP...08..026K}; we find no pulsars with a $\sim\mathrm{ms}$ birth period in our sample. This is consistent with some previous studies utilising archival X-ray data \citep{10.1111/j.1365-2966.2007.12821.x}.

\begin{figure}
\centering
    \includegraphics[width=\columnwidth]{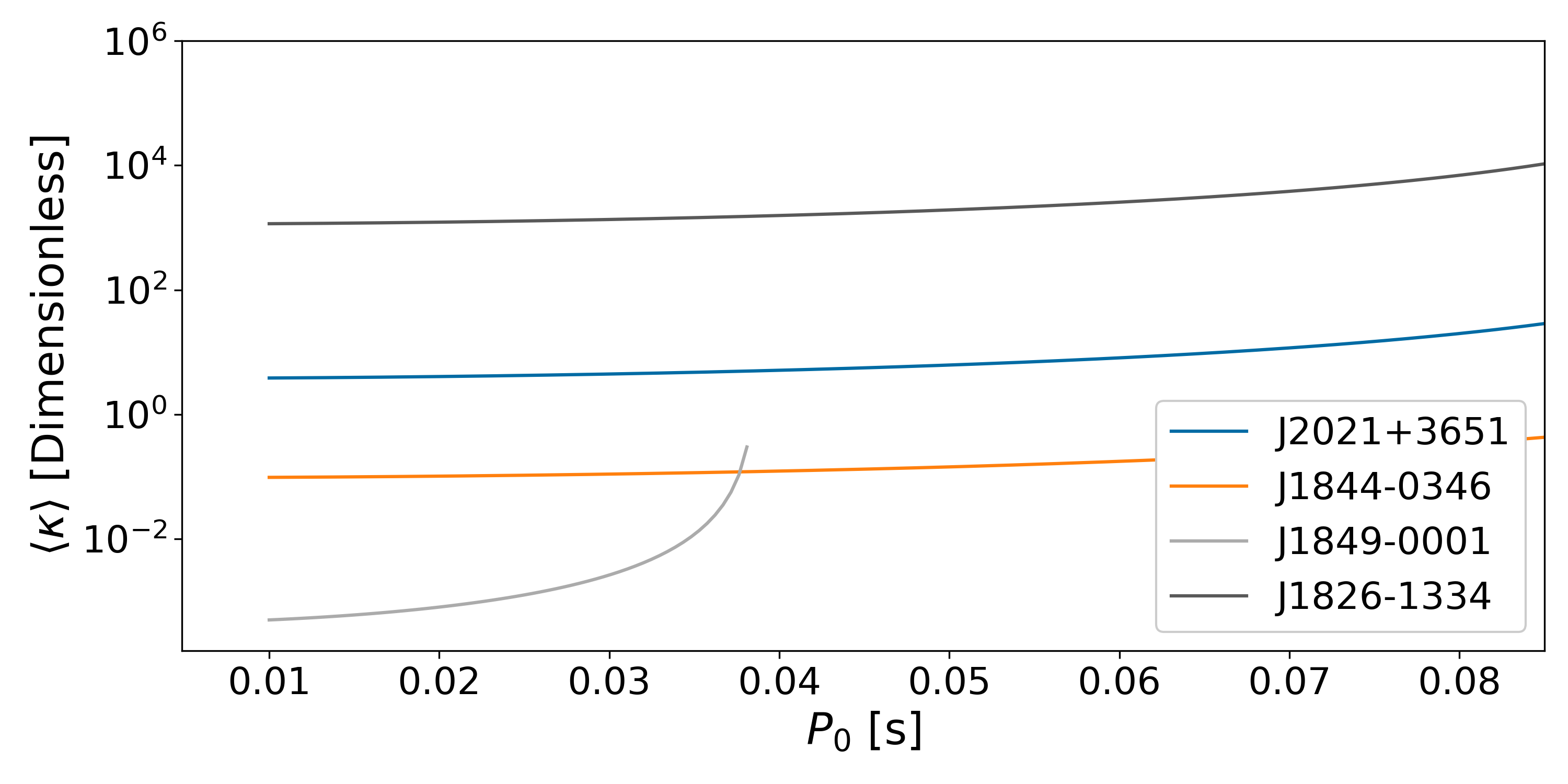}
    \caption{Calculated lower limits on the pair-production multiplicity for the four pulsars co-incident with LHAASO sources.}
    \label{fig:lhaasoplot}
\end{figure}

The results of the modelling of the four LHAASO sources can be seen in Fig. \ref{fig:lhaasoplot}. We cannot attempt to derive values for $\langle\kappa\rangle_{\rm lim}$ as we do not have radio observations for these targets. However, if we assume the potential range of $\langle\kappa\rangle_{\rm lim}$ values for these systems is comparable to those we have determined with the H.E.S.S. data, we cannot exclude the possibility of hadronic particles reaching the pulsar wind for any of these systems either.

\begin{table*}
\centering
\caption{Derived values of $P_0$, along with $N_{el}$, the total energy in electrons $E_{\mathrm{e,tot}}$, and the lower limit for $\langle\kappa\rangle$ for the six H.E.S.S. sources where this is possible. }
\begin{tabular}{c c c c c c}
\hline
\hline
\makecell{ATNF Name\\} & \makecell{$P_0$\\$\mathrm{[ms]}$} & \makecell{$\dot{E}_t$ \\$\mathrm{[\times 10^{36}\ erg/s]}$} & \makecell{$N_{el}$ \\$\mathrm{[\times 10^{47}\ Counts]}$} &\makecell{$E_{\mathrm{e,tot}}$\\$\mathrm{[\times 10^{47}\ erg]}$}&\makecell{$\langle\kappa\rangle$\\$\mathrm{[Dimensionless]}$}\\
\hline
J1833-1034 &  33.0 & $33.9$& $45.6$ &$51.8$& 1476\\
J1513-5908 &   15.2 & $17.0$& $5.14$ & $8.35$&809\\
J1930+1852 &  41.3 & $12.0$ & $5.06$ &$11.0$&432\\
J1846-0258 & 50.8 &$8.13$ & $1.62$ & $1.87$& 1061\\
J0835-4510 &  10.9 & $6.92$ & $18.7$ & $24.7$&174\\
J1747-2809 & 47.9 &$42.7$ & $125$ &$71.4$ & 29328\\
\hline
\end{tabular}
\tablefoot{$\langle\kappa\rangle$ lower limit values obtained by finding the intersection between the line of $\langle\kappa\rangle(P_0)$ and our derived $P_0$ in Fig. \ref{fig:kappaplotdual}.}
\label{table:kappatab}
\end{table*}
\section{Discussion}
\label{sec:discussion}
\subsection{Implications for production of cosmic rays}
We cannot safely exclude the possibility of hadrons escaping the pulsar surface into the wind for any of our considered sources. However, it should be noted that the values of $\langle\kappa\rangle$ we observe for all sources are within likely theoretical constraints (given $P_0$ is not likely to be $\gg$50ms), which cap the maximum pair production multiplicity to be a few times $10^5$ \citep{Timokhin_2019}. The $\langle\kappa\rangle$ curve we observe for PSR J1849-0001 is significantly below unity, but similar findings were encountered by \citet{de_Jager_2007}, who found that the Vela Pulsar (PSR J0835-4510) had a pair production multiplicity less than unity (depending on the assumptions made).

\subsection{The phase-space of free parameters for the H.E.S.S. sources}
\begin{figure}
\centering
    \includegraphics[width=\columnwidth]{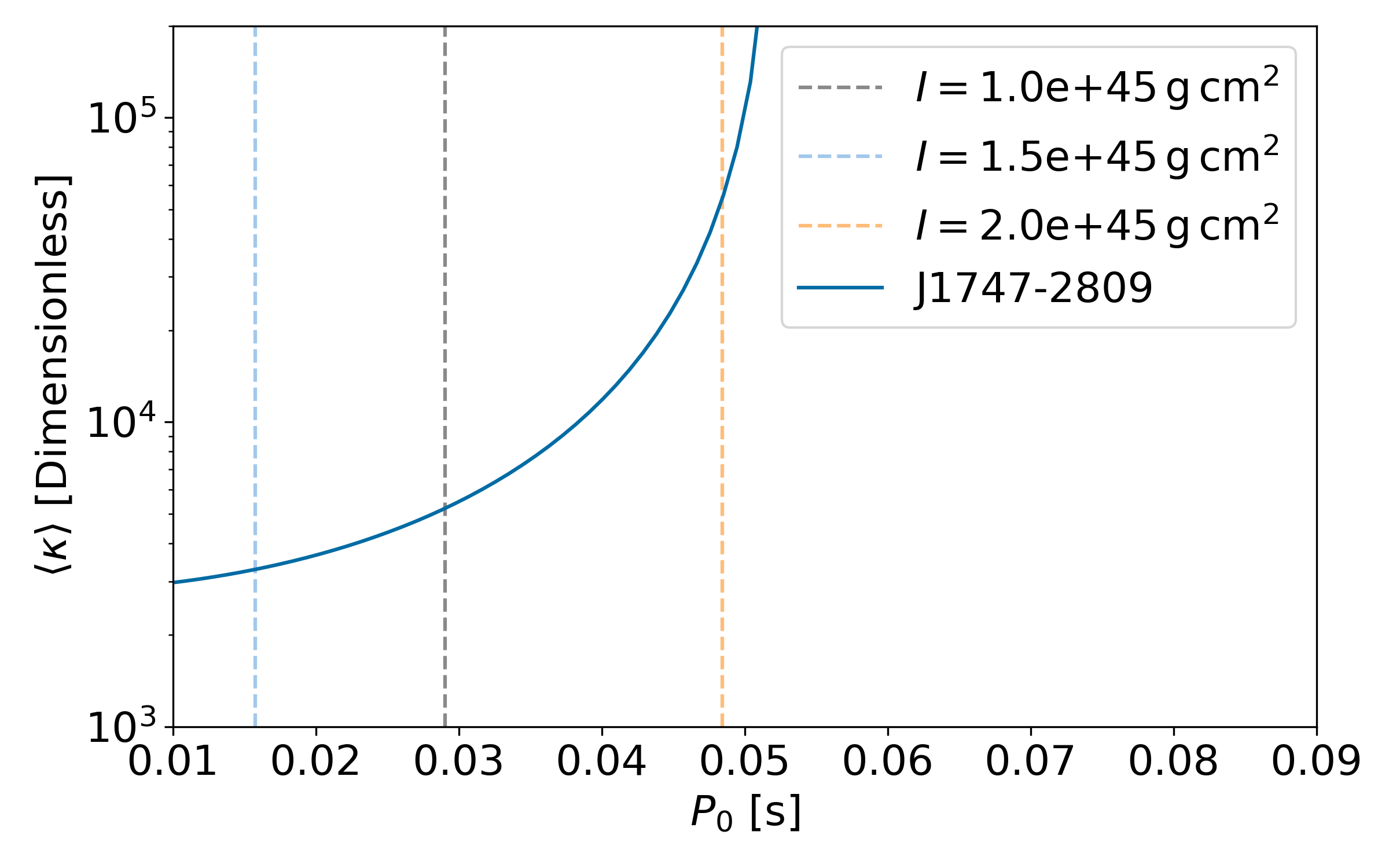}
    \includegraphics[width=\columnwidth]{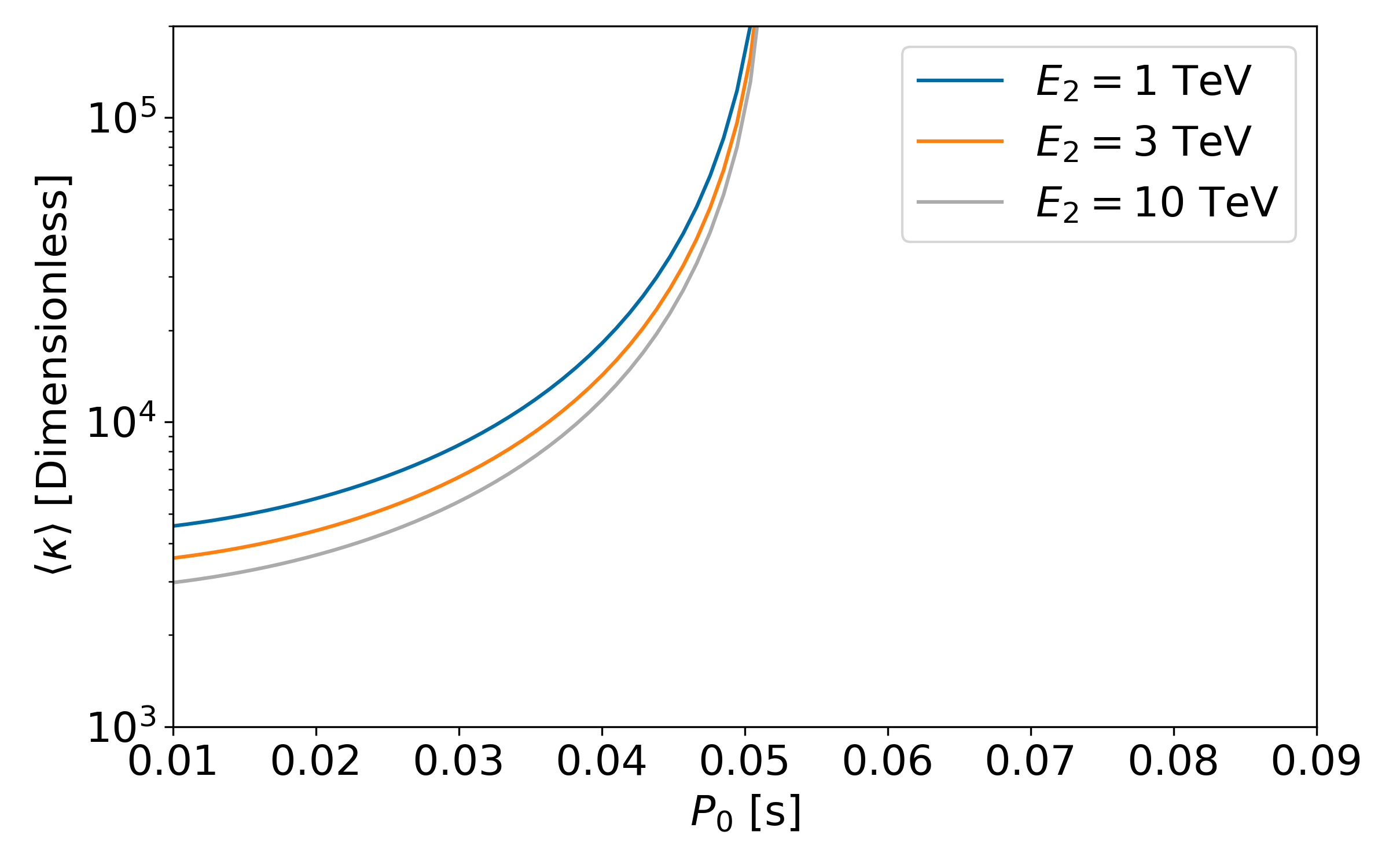}
    \includegraphics[width=\columnwidth]{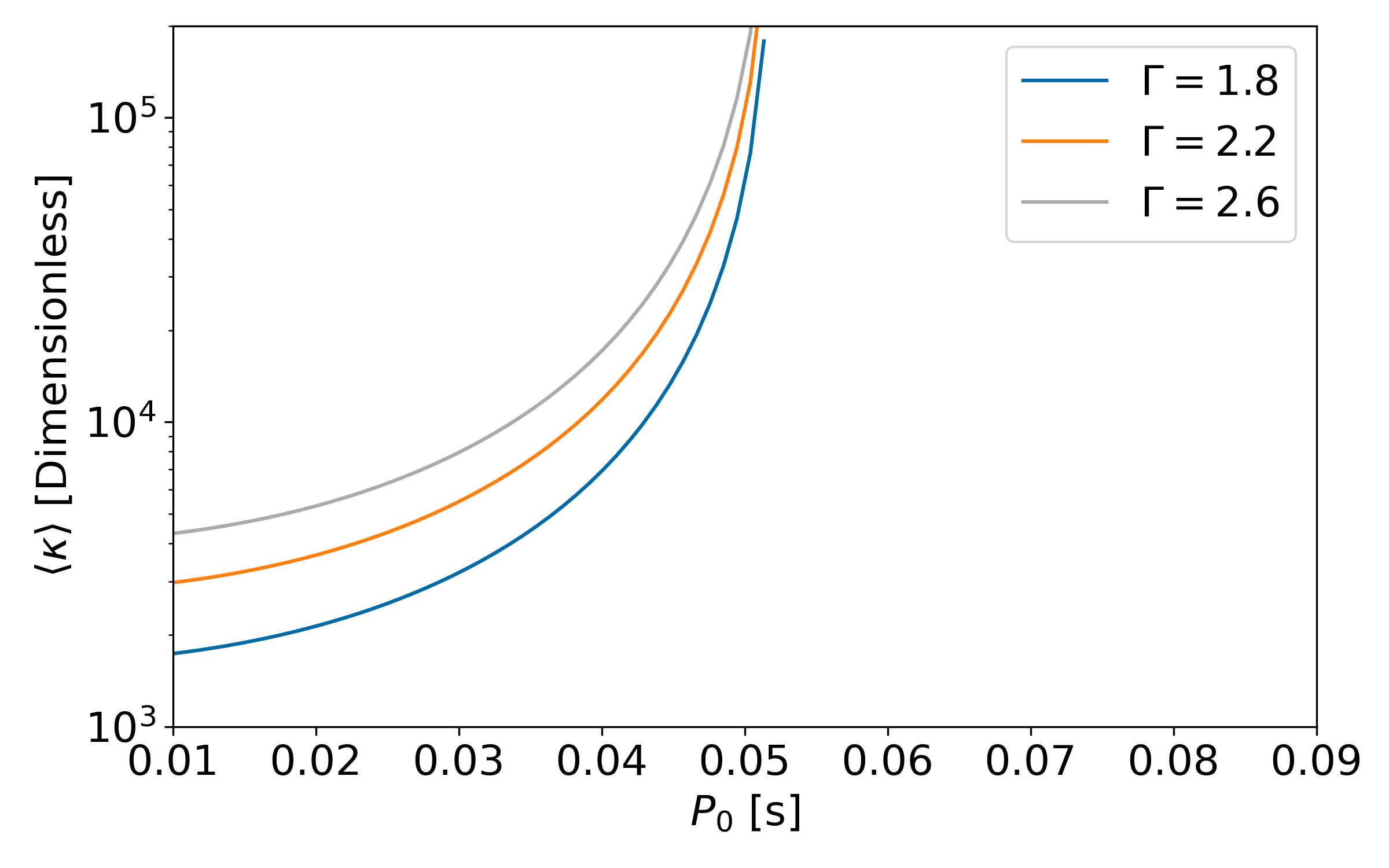}
    \includegraphics[width=\columnwidth]{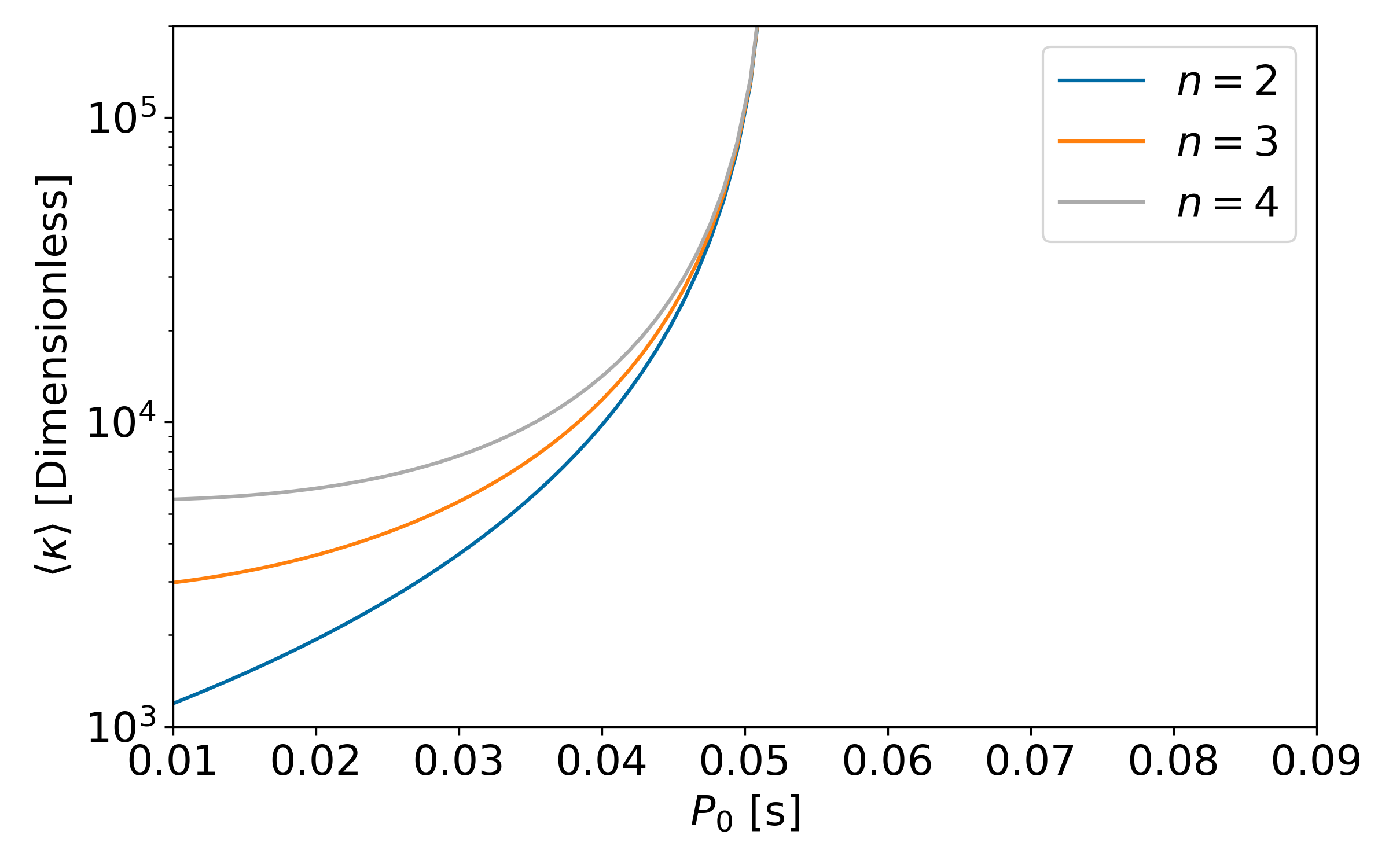}
    \caption{Effect of varying $I$ (top), $E_2$ (upper middle), $\Gamma$ (lower middle), and $n$ (bottom) for PSR J1747-2809, which has the lowest $P_t$ (47.9 ms) in our sample of pulsars with a constrained $P_0$.}
    \label{fig:gamscan2}
\end{figure}

\begin{figure}
\centering
    \includegraphics[width=\columnwidth]{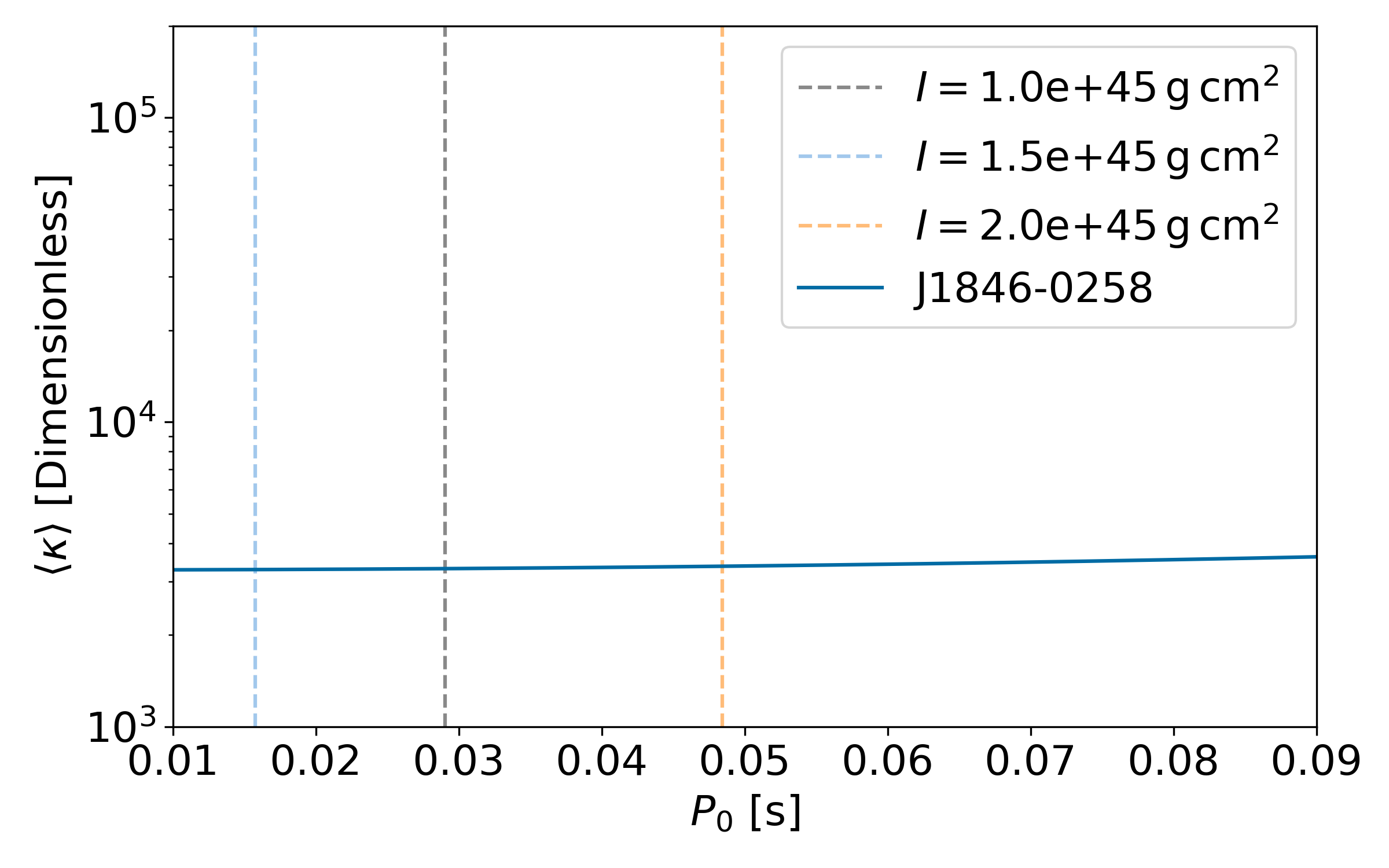}
    \includegraphics[width=\columnwidth]{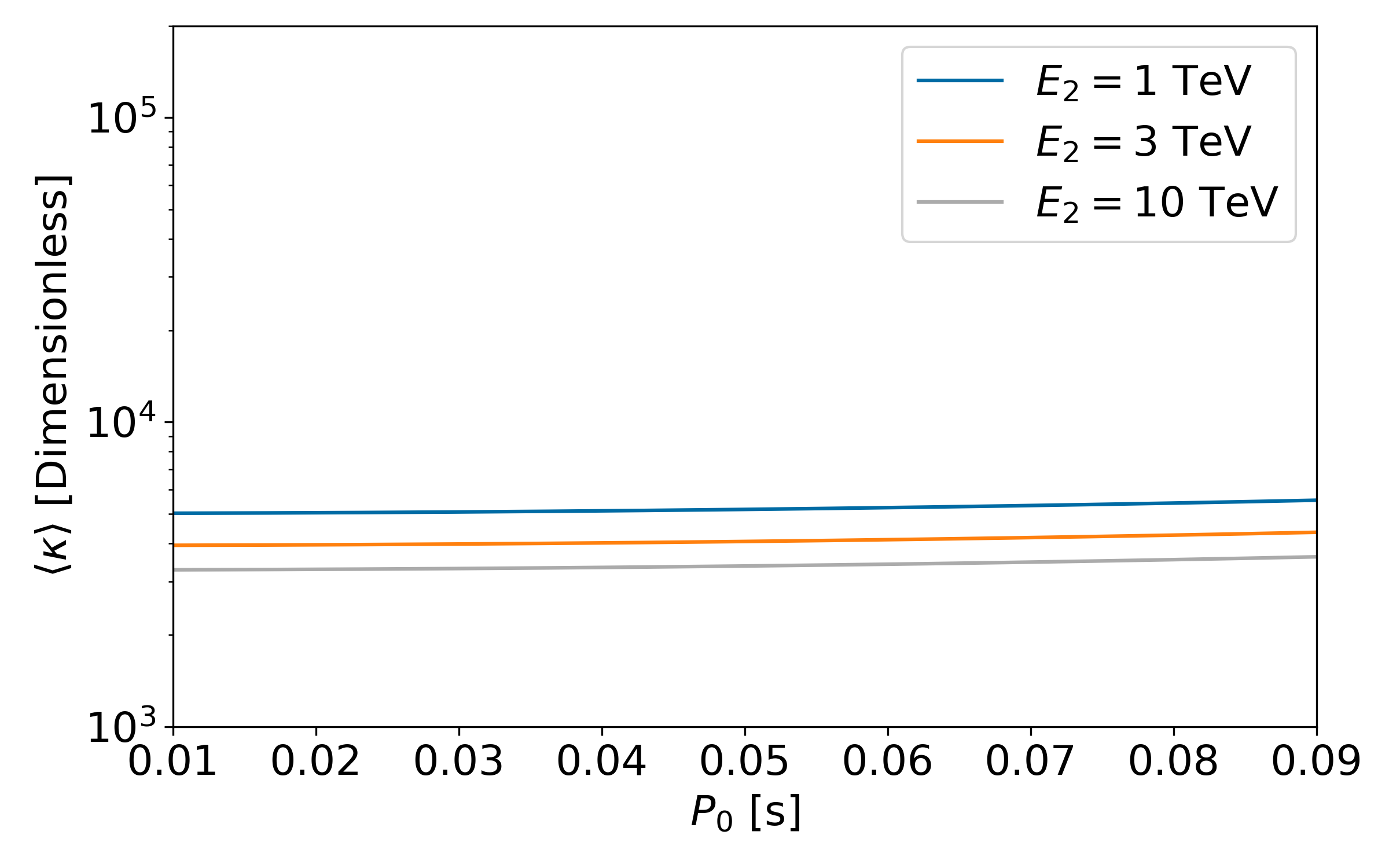}
    \includegraphics[width=\columnwidth]{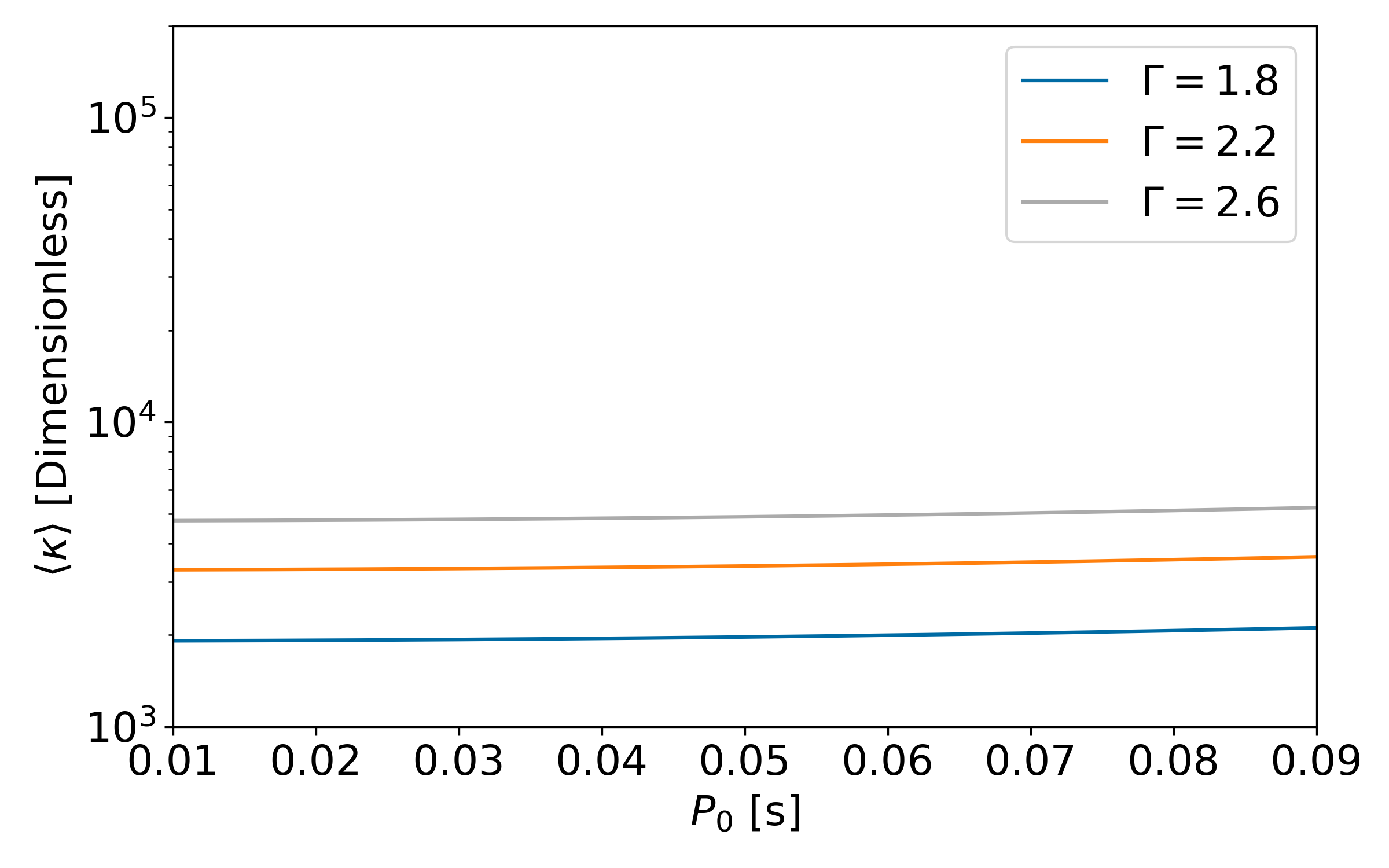}
    \includegraphics[width=\columnwidth]{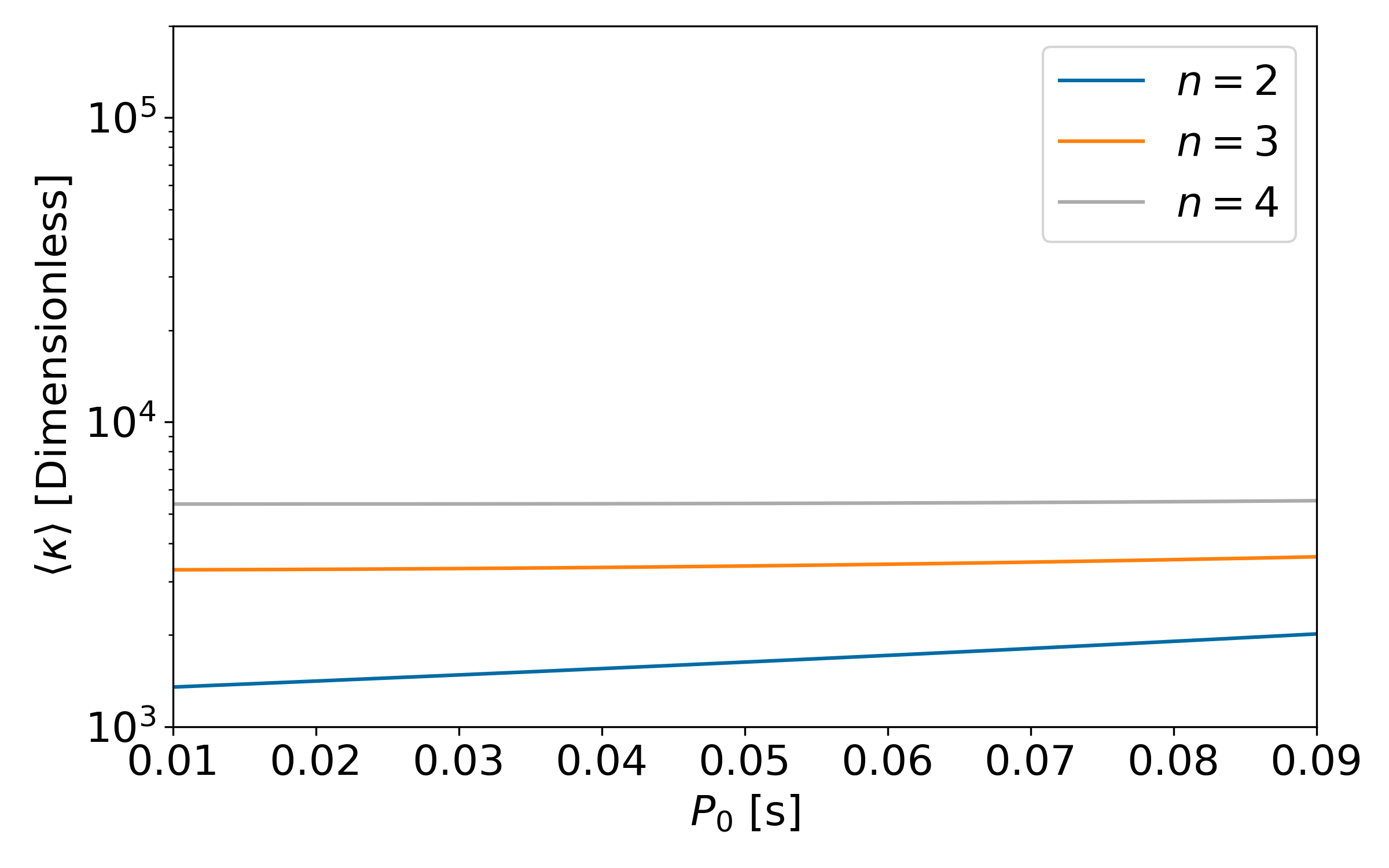}
    \caption{Effect of varying $I$ (top), $E_2$ (upper middle), $\Gamma$ (lower middle) and $n$ (bottom) for PSR J1846-0258, which has the highest $P_t$ (326.6 ms) in our sample of pulsars with a constrained $P_0$.}
    \label{fig:gamscan}
\end{figure}

As our work is based on the previous findings of \citet{Giacinti_2020}, which did not provide estimates of the uncertainties on the PWN and SNR radii, we cannot perform a full, robust uncertainty analysis. Instead, to investigate the effect of various assumptions in our models, we have examined the effect of changing parameters in our model for the two pulsars PSR J1747-2809 and PSR J1846-0258 (to probe the minimum and maximum values of $P_t$ respectively). The results of this are shown in Figs. \ref{fig:gamscan2} and \ref{fig:gamscan}. 

The selected value of $I$ appears to have an approximately order of magnitude effect on the point of intersection between the pair production multiplicity curve and the pulsar birth period estimate for the short period pulsar PSR J1747-2809 (seen in Fig. \ref{fig:gamscan2}); however, it exerts very little influence upon the derived pair-production multiplicity, as seen in Fig. \ref{fig:gamscan} for PSR J1846-0258. Variations in the moment of inertia $I$ correspond to differences in the internal structure and can be used to constrain the equation of state for dense matter \cite{1994ApJ...424..846Ravenhall}. The results for both pulsars suggest there is an impact on the derived $P_0$ by roughly a factor of 2, within the band of reasonable potential values. Changing the value of $E_2$ is seen to only have a modest effect on the value of $\langle\kappa\rangle$ in Figs. \ref{fig:gamscan2} and \ref{fig:gamscan}. Further increasing $E_2$ would lead to a resulting decrease in $\langle\kappa\rangle$. Choosing 10 TeV for this value appears to be reasonable as a lower limit given our relative agreement for the Vela pulsar and PSR B1509-58 with \citet{de_Jager_2007}, where the pair-production multiplicity is calculated differently.  Changing the value of $\Gamma$ by 0.4 appears to have an approximately factor two effect on the derived $\langle\kappa\rangle$. The true pair production multiplicity value is very likely to be within this range, as a value of 2.2 is expected for particles at an ultra-relativistic shock undergoing isotropic scattering \citep{10.1046/j.1365-8711.2001.04851.x}. A similar magnitude effect is seen when varying $n$ for both pulsars and this is likely to dominate the uncertainty for shorter birth period pulsars. However, as this braking index has only been measured for a very small population of pulsars, by convention, we adopted a value of 3, previously adopted by many works in the literature as well \citep{gaensler_slane_review}.

\subsection{Prospects for future observations}

The large beaming fraction (up to 0.92) for gamma-rays from high $\dot{E}_t$ pulsars \citep{aris} suggests that observations of gamma-ray pulsars are nearly complete. This implies conclusions of this study regarding the contribution of pulsars to the hadronic cosmic ray spectra are unlikely to change with the advent of future gamma-ray instruments, such as the Cherenkov Telescope Array \citep{science_with_CTA} or the Southern Wide-Field Gamma-Ray Observatory \citep{SGSO_science_case}. That said, state-of-the-art and future radio observations by observatories such as MeerKAT \citep{goedhart2023sarao} and the upcoming Square Kilometer Array \citep{Carilli_2004} could allow for more PWN and SNR radii, from systems with lower surface brightness and larger extent, that are yet to be measured. This means the pair production multiplicities and birth period could potentially be derived for more pulsars. In combination with multi-wavelength modelling, this would allow for more stringent constraints on hadronic contributions to the gamma-ray emission from PWNe.


\section{Conclusions}
\label{sec:conclusions}
In this study, we have placed constraints on the pair production multiplicity for 26 pulsars observed by H.E.S.S., deriving lower limits for the average pair production multiplicities and birth periods for six of these pulsars. We also set lower limits on the pair production multiplicities for four pulsars that are co-incident with sources observed by LHAASO, based on their very-high-energy gamma-ray emission. We find that the possibility of hadrons escaping the pulsar into the wind cannot be conclusively excluded for any source we consider here. Nevertheless, for J1747-2809 (the pulsar for which we infer the highest pair-production multiplicity), the escape of hadrons into the wind is essentially ruled out for values of $\eta \geq 0.1$. This implies that hadrons are only released into the wind of J1747-2809 if $\eta \leq 0.1$, which is too low for any hadronic component to be energetically significant. %

\begin{acknowledgements}
This work is supported by the Deutsche Forschungsgemeinschaft (DFG, German Research Foundation) – Project Number 452934793.
\end{acknowledgements}

\bibliographystyle{aa}
\bibliography{references.bib}

\end{document}